# Metalens array for complex-valued optical discrete Fourier transform

Randy Stefan Tanuwijaya[1], So Lap[1], Wai Chun Wong[1], Tailin An[1],

Wing Yim Tam[1], and Jensen Li[1,2]

[1]Department of Physics, The Hong Kong University of Science and Technology,
Clear Water Bay, Kowloon, Hong Kong, China

[2]Department of Engineering, University of Exeter, Exeter, United Kingdom

Photonic computing has emerged as a promising platform for accelerating computational tasks with high degrees of parallelism, such as image processing and neural network. We present meta-DFT (discrete Fourier transform), a single layer metasurface device, designed to perform optical complex-to-complex DFT with $\mathcal{O}(N)$ time complexity. One critical challenge in free-space analog optical computing is to control the measurement error. Our scheme addresses this issue by focusing light on spatially separated focal points and reconstructing the complex phase, which enable error correction. We systematically evaluate the device's performance using input vectors with random complex amplitudes and phases, to demonstrate its robust accuracy. Our findings pave the way towards advancement of metasurface-based computation, offering a robust framework that is readily extensible to an arbitrary complex-valued matrix-vector multiplication (MVM).

**Introduction**

The growing demand in machine learning and artificial intelligence models has brought focus into research of more energy efficient alternatives to traditional electronic computing frameworks [1]. Photonic computing emerges as a promising platform, offering advantages such as higher speed, larger bandwidth, lower power consumption, and enhanced parallelization capabilities [2,3,4]. Due to their inherent ability to achieve parallelization at the physical level, photonic computers excel in tasks requiring high degrees of parallelism, including signal/image processing [5,6,7,8], solving numerical equations [9,10,11], neural network [12,13,14,15,16,17], and quantum computing [18,19,20].

Traditionally, photonic computing can be implemented by common optical elements like lens, mirrors, and waveplates. For photonic computers to serve as viable accelerators for computing tasks, miniaturization is essential for seamless integration into existing infrastructures, along with robust error control mechanisms. Recent advancements in nanofabrication technologies have enabled the development of nanophotonic devices, such as metasurfaces and photonic integrated circuits. Metasurfaces – two-dimensional arrays of nanostructures, is a promising platform for free-space photonic computing. This approach offers higher bandwidth and greater parallelism compared to on-chip methods [21]. The large design degree-of-freedom, compact form factor, and cost-effective fabrication of metasurfaces have led to successful applications, including beam steering [22,23], generation of vortex beam [24,25,26], multi-channel hologram generation [27,28,29,30,31,32], and numerous computing applications [4].

While computing in free space provides a larger information bandwidth, it is often susceptible to errors due to misalignment of the optical elements. In contrast, photonic integrated circuits localize each input/output field within waveguides, creating discretized channels for computation that minimizes errors. Although metasurface platforms operate in free space, they can be engineered to localize fields and generate discretized output fields. Localizing the field embeds redundancy of information and allows error correction, as demonstrated in the optical neural network (ONN) framework [13] and simulation of quantum algorithm [33]. Central to these photonic computing demonstrations is wavefront manipulation, which can be mathematically modelled as complex matrix-vector-multiplication (MVM). However, output measurements typically only capture the intensity of the electric field, resulting in the loss of complex phase information. While interferometric methods can recover this phase, they often necessitate additional optical elements or computations [34,35].

In this paper, we introduce meta-DFT—a single-layer metasurface platform for optical complex-to-complex discrete Fourier transform (DFT). While we use DFT as an example, our design can be generalized to other complex matrices. By leveraging optical parallelism, we achieve $N \times N$ complex-to-complex MVM that scales with $\mathcal{O}(N)$ rather than $\mathcal{O}(N^2)$. Our design features an array of metalens to generate multiple focal points, eliminating the need for a separate reference beam to recover the complex phase. Additionally, we also develop an error correction scheme by analyzing localized interference patterns at each focal point. We experimentally demonstrate the device's functionality, error correction scheme, and quantitative performance at near-infrared (NIR) wavelength. Finally, we highlight our platform's versatility for arbitrary complex-valued matrix-vector multiplication and its scalability for larger matrices.

# Method and device

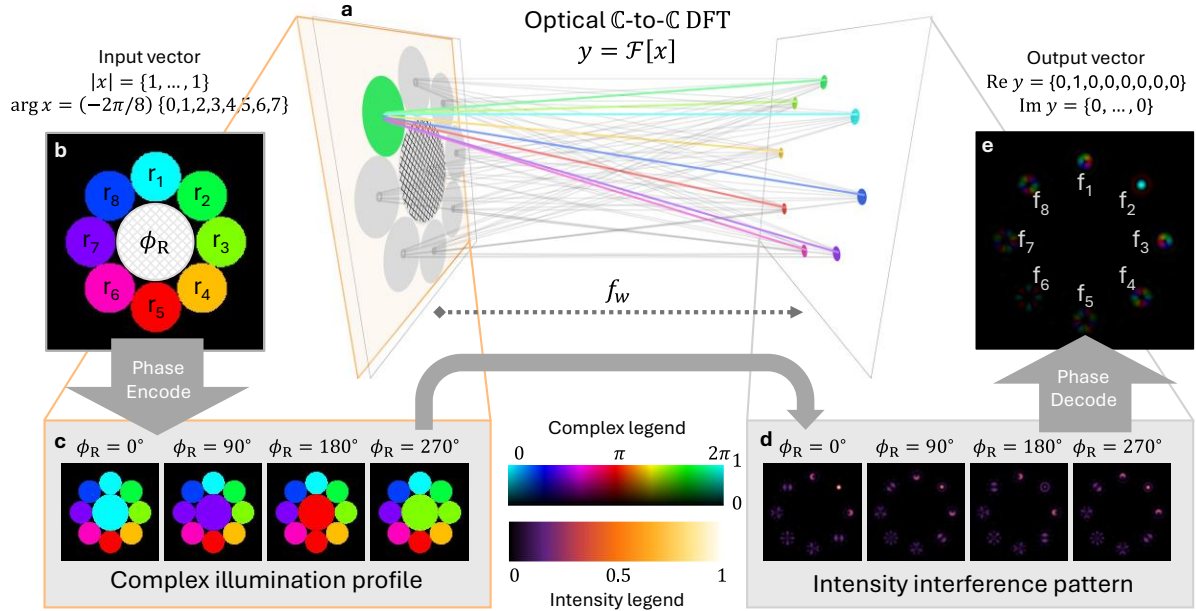

**Figure 1. Schematics of meta-DFT scheme. (a)** The eight circumferential metalenses focus light into 8 focal points to achieve complex-to-complex DFT. The central reference metalens (at the center) is used to recover the phase information. Each metalens is designed to focus light on 8 focal points, represented by the lines connecting them. The colored lines (all at the same wavelength) illustrate the light focused from the metalens at $r_2$ to every focal point. The hue indicates the complex phase of the focused light, which differ by $2\pi/8$ for neighboring focal point. Interference patterns emerge at each focal point due to the interactions of light from different metalenses. **(b)** An example of input vector $x$, with $l_x = -1$, encoded as metasurface complex illumination profile. **(c, d)** The illumination profile and corresponding intensity interference pattern by iterating four different reference lens phases $\phi_R = \{0°, 90°, 180°, 270°\}$. **(e)** The reconstructed field by combining the four intensity interference patterns. The output vector can be obtained by reading the complex value of the pixel at the center of each complex interference pattern,

Our proposed meta-DFT scheme utilizes an array of metalenses to focus light into multiple focal points, as schematically illustrated in Fig. 1a. The primary objective is to perform an N-dimensional DFT of an input vector $x = \{x_1, ..., x_N\}$ to yield output vector $y = (\text{DFT}_N)x = \{y_1, ... y_N\}$. We design $N = 8$ signal metalenses to optically execute discrete Fourier transform. The signal metalenses are arranged in a circular configuration on the left plane of Fig. 1a, with each metalens centered at $r_1$ to $r_8$ being a basis of the input vector. An additional reference metalens are added at the center for phase retrieval. Each metalens is designed to focus light on 8 focal points at $f_1$ to $f_8$ on the focal plane, with focal length $f_w$, as shown on the right plane of Fig. 1a.

The input vector $x$ is encoded by modulating the illumination amplitude and phase profile of each metalens. This can be done by using a spatial light modulator (SLM) [36], to shine light with a given amplitude and phase onto each metalens. For example, $x_j$ can be encoded as the illumination profile of the signal metalens at $r_j$. In Fig. 1b, we show the complex illumination profile for a sample input vector $x$; with $\arg x = -\frac{2\pi}{8}\{1,2,\ldots,8\}$ and all elements having unit amplitude. The black region indicates area of no illumination. When this input vector is encoded as illumination profile, the incident light onto the metalens yields a topological charge of $l_x = -1$ for this example, as the phase wraps $2\pi$ once in the counterclockwise direction.

To implement the optical DFT, we design how much light is to be transferred across the metasurface and focused onto each focal point and the phase delay associated with it, according to the design matrix $T$:

$$T_{ij} = \begin{cases} \frac{1}{8}\exp\left(i\frac{2\pi}{8}(i-1)(j-1)\right) & \text{(singal: } 1 \leq i,j \leq 8\text{)} \\ \frac{1}{8} & \text{(ref: } 1 \leq i \leq 8, j = 9\text{)} \end{cases} \quad (1)$$

The first eight columns of this matrix ($1 \leq j \leq 8$) correspond to the 8 signal lenses and implements $T_{ij} = (\text{DFT}_8)_{ij}$. Each signal metalens will focus light into each focal point with equal amplitude and varying phase delay. The last column ($j = 9$) corresponds to the reference lens, which focuses light onto the eight focal points with same amplitude and phase.

In Fig. 1a, the lines connecting every metalens with every focal point represent one element in the $8 \times 9$ design matrix $T$. The colored lines denote light from the signal metalens at $r_2$, and their hues indicate the phase at each focal point $f_i$, $\arg(T_{i2}x_2)$ – for illumination phase $\arg x_2 = -2\pi/8$. The color variation reflects the $2\pi/8$ phase difference between neighboring focal points, with light focused to $f_2$ yielding a phase of 0, as indicated by the cyan color

At the focal plane of the metalens, the interference patterns form at each focal point $f_i$ and are captured by a camera. Even without any illumination on the reference lens, it is possible to obtain the complex amplitude of the output vector $|y|$, by measuring the intensity at the center of each focal point [33]. To retrieve the output complex phase, we combine the interference patterns from experiment with varying illumination phase on the reference lens. The light from the reference lens perturb the signal, causing the intensity at each pixel to vary depending on the relative phase between the signal and reference metalenses.

As shown in Fig. 1c, we iterate four different phases on the reference lens $\phi_R = \{0°, 90°, 180°, 270°\}$. Each plot illustrates the complex illumination field $E_{sig}(x) + E_{ref}(\phi_R)$ on the metasurface plane. The black region indicating no illumination. By capturing an image on the focal plane of the metasurface, we obtain four intensity images $I(x; \phi_R = 0°)$, $I(x; \phi_R = 90°)$, $I(x; \phi_R = 180°)$, and $I(x; \phi_R = 270°)$, as shown in Fig. 1d. Here, $I(x; \phi_R) = |\mathcal{F}[E_{sig}(x) + E_{ref}(\phi_R)]|^2$ represents the intensity profile at the focal plane for a given input vector $x$ and reference lens phase $\phi_R$; $E_{sig}$ and $E_{ref}$ denote the fields of the signal and reference metalens at the metasurface plane, while $\mathcal{F}$ denote propagation of the field from the metasurface plane to the focal plane, which can be simulated by angular spectrum method. The reconstructed complex field $E(x)$ can be derived using the following equations:

$$\begin{aligned} \text{Re}[E(x)] &= I(x; \phi_R = 0°) - I(x; \phi_R = 180°) \\ \text{Im}[E(x)] &= I(x; \phi_R = 90°) - I(x; \phi_R = 270°) \end{aligned} \quad (2)$$

Notably, $E(x) \propto \mathcal{F}[E_{sig}(x)] \mathcal{F}[E_{ref}(\phi_R = 0°)]^*$ represents the reconstructed field at the focal plane. Fig. 1e illustrates the reconstructed field on the focal plane $E(x)$ for input vector $x$ with $l_x = -1$. The brightness of each pixel shows the complex amplitude, while the hue shows the complex phase. In this example, applying DFT onto the input vector $x$ yields $y = \{0,1,0,0,0,0,0,0\}$. There is only a single nonzero element in $y$ since the input vector consists solely of a single OAM mode. This corresponds to a constructive interference pattern at the second focal point $f_2$ and destructive patterns at others. For constructive interference pattern at $f_2$, we obtain the complex phase of 0 radian, as indicated by the cyan color.

In essence, the focal point at $f_2$ acts as as a detector for input topological charge $l_x = -1$. This is due to the entire metalens array performing the DFT by focusing incident light to each focal point $f_i$ with added topological charge of $(i - 1)$. For example, the topological charge $l_y$ of the reconstructed field at focal point $f_i$ is given by $l_y(f_i) = l_x \oplus (i - 1)$ (Where $\oplus$ denotes cyclic addition constrained from $-3$ to $\pm 4$). During propagation, the topological charge is conserved, leading to the observation of orbital angular momentum (OAM) beam structures at the focal points.

For instance, destructive interference patterns at $f_1$ and $f_3$ exhibit OAM beam structure with $l_y = -1$ and $+1$, respectively.

To realize our meta-DFT device, we designed a transmissive type metalens array with the following profile:

$$t_j(\rho_u, \rho_v) = \frac{1}{A}\sum_i T_{ij} \exp\left(-i\frac{2\pi}{\lambda}\sqrt{|\rho_u - f_{iu}|^2 + |\rho_v - f_{iv}|^2 + f_w^2}\right) \quad (3)$$

The equation specifies the design transmission amplitude and phase for the unit cell located at $(\rho_u, \rho_v)$ (global coordinate measured from the center of the entire metasurface) within the aperture of metalens $j$. The subscripts, $(u, v, w)$ denote the coordinate in the 3D space. The design focal length $f_w$ is the same across all focal points, while $(f_{iu}, f_{iv})$ denotes the transverse position of the focal point $f_i$. Note that $t_j$ is the desired transmission profile, while the actual transmission also depends on transmission efficiency of the unit cell. Each column and row of this design DFT matrix corresponds to a specific metalens and focal point, correspondingly. $A$ is normalization factor, to make sure $|t_j| \leq 1$ for practical realization. Furthermore, the design allows flexibility to implement general matrices by modifying the design matrix $T$.

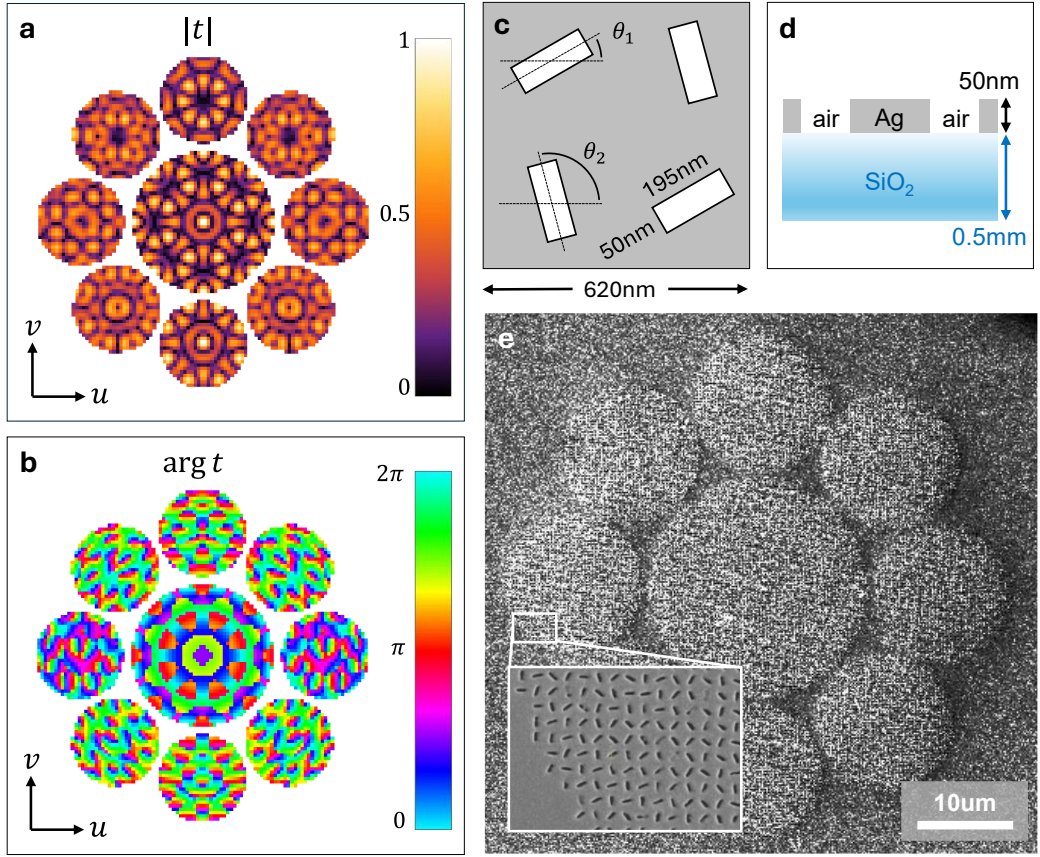

**Figure 2 Metasurface sample for meta-DFT. (a, b)** Transmission amplitude and phase profiles. **(c)** 2x2 geometric phase unit cell design for operational wavelength at $\lambda = 810$nm (620nm periodicity). The cross-circular polarization transmission amplitude and phase are controlled by adjusting the rotation angles $\theta_1$ and $\theta_2$ of the two nanoslots. **(d)** Cross section of the metasurface sample (not to scale). **(e)** SEM image of fabricated metalens array sample.

The metasurface sample for meta-DFT is shown on Fig. 2. Fig. 2a and 2b illustrate the amplitude and phase of the metalens array profile. The metalenses are designed with a focal length $f_w = 120$μm, and $f_i$ positioned directly behind the center of each metalens. The signal and reference metalenses consist of 26 and 42 unit-cells along the diameter, respectively. The unit-cell periodicity is 620nm, and the overall dimensions of the metasurface is 60μm × 60μm.

To translate this transmission profile into our metasurface design, we employ a 2x2 geometric phase unit cell configuration, as shown in Fig. 2c. Each slot is optimized as a polarizer at 808nm wavelength, with cross-circular polarization conversion efficiency of around 15%. By adjusting the rotation angles of the two slots, we can control both the cross-circular polarization transmission amplitude and phase for each unit cell. The difference of the two rotation angles determines the transmission amplitude $|t(\rho)| \propto |\cos(\theta_1 - \theta_2)|$, while the sum controls the transmission geometric phase $\arg t(\rho) = \theta_1 + \theta_2$. It is important to note that the amplitude control accompanies loss through absorption. Fig. 2d presents the cross-section of the sample, fabricated using the focused ion beam (FIB) technique, which features 50nm thick silver layer on glass. The scanning electron microscopy (SEM) image of the fabricated sample is shown in Fig. 2e.

**Results and performance**

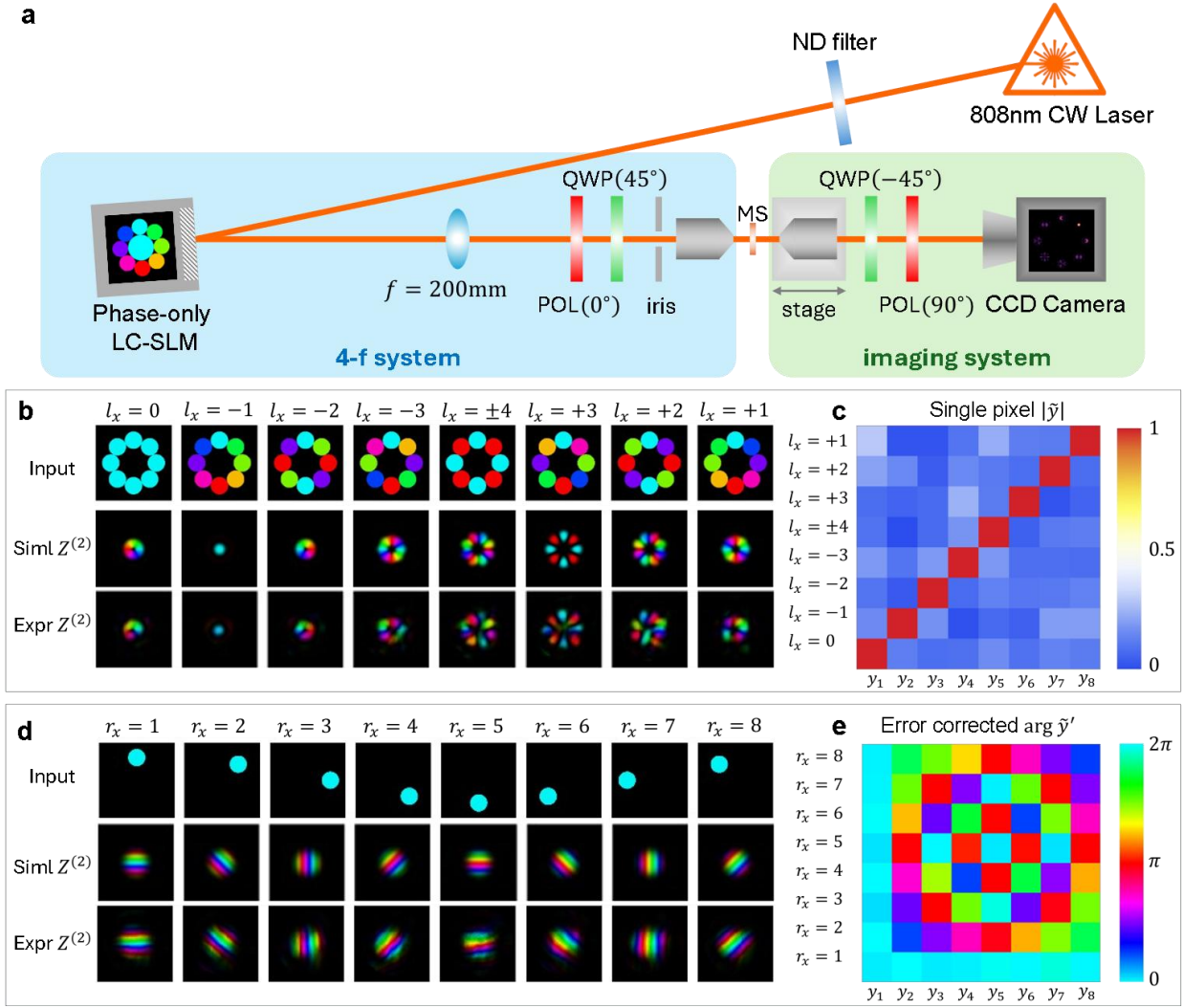

Figure 3 Setup and calibration experiment. (a) Experimental setup. (b) Calibration/OAM input experiment. From top-to-bottom rows: SLM profiles, simulated and measured reconstructed fields around $f_2$. (c) Measured normalized complex amplitude of the center pixel $|\tilde{y}|$. (d) Single lens illumination experiment, with same format as (b). (e) Measured complex phase $\arg \tilde{y}$ after error correction.

The experimental setup for characterizing the meta-DFT device is presented in Fig. 3a. An 808nm continuous-wave laser serves as a coherent light source. The input vector is encoded using a phase-only SLM, which is imaged onto the metasurface using a 4-f system composed of a $f = 20$cm lens and a 20x objective lens. A phase grating is employed to control the amplitude by diffracting light away from the metasurface (phase grating appears as black region in the illumination profile). Light diverted by the phase grating will be blocked by the iris positioned at the focal plane of the $f = 20$cm lens.

A combination of polarizers and quarter wave plates are used to illuminate metasurface with LCP polarized light, and to observe the interference pattern with RCP polarization. Another 20x objective mounted on a translational stage is used to image on the metasurface plane (for alignment) or the focal plane (for measurement) onto a CCD camera.

Fig. 3b shows calibration/OAM input experiment. The first row shows SLM illumination profiles, with $l_x \in L = \{0, -1, -2, -3, \pm4, +3, +2, +1\}$. The second and third row presents the simulated and experimental complex field at $f_2$. An additional topological charge of $+1$ is added to this focal point, which shows a constructive interference when the input beam $l_x = -1$. The output vector is measured by reading the value of a single pixel at the center of each complex interference pattern, expressed as: $\tilde{y}_i = \frac{1}{A_i} E^{(i)}_{\{u_0,v_0\}}(x)$. Here, $E^{(i)}_{\{u_0,v_0\}}$ denotes the complex value of the pixel at $\{u_0, v_0\}$ at the center of the complex interference pattern at $f_i$, and $A_i$ is the maximum measured value, i.e. measured value at constructive interference when $l_y(f_i) = 0$.

Fig. 3c shows the measured amplitude of $|\tilde{y}_i|$ obtained from the single pixel measurements. Ideally, the amplitudes should be 1 for the diagonal elements and 0 for the off-diagonal elements. The root-mean-square error (RMSE) for the measured output vector amplitudes $|\tilde{y}|$ in this experiment is 0.12. In principle, we can obtain the complex amplitude and phase by reading the complex value of a single pixel. On the other hand, since information from the surrounding pixels is highly correlated with the center pixel, the surrounding pixels can be useful to formulate an error correction method for our meta-DFT scheme.

To perform the error correction, we use the eight OAM input beam shown in Fig. 3b as part of the calibration process. The eight reconstructed fields obtained for a particular focal point $f_i$, can be used to construct a complex mask $m^{(i)}$ by using a pseudoinverse method, adopting a compressive sensing approach. We can use $K$ pixels around each focal point to construct the mask, where $1 \leq K \leq 80^2$, where $80 \times 80$ is the dimension of the experimental images for each interference pattern at $f_i$. The error-corrected output vector $\tilde{y}$ can be computed as the dot product between the reconstructed field around $f_i$, $E^{(i)}(x)$ and the mask $m^{(i)}$:

$$m^{(i)} = \left[E_{\text{cali}}^{(i)}\right]^{-1}\left[y_{\text{cali}}^{(i)}\right] \implies \tilde{y}_i = \sum_{u,v} m_{\{u,v\}}^{(i)} E_{\{u,v\}}^{(i)}(x) \tag{4}$$

Here, $\left[E_{\text{cali}}^{(i)}\right]$ is a $8 \times K$ matrix obtained from the calibration experiment, and $[y_{\text{cali}}^{(i)}]$ is $8 \times 1$ target calibration values of $y_i$. We adopt Moore-Penrose pseudoinverse as $\left[E_{\text{cali}}^{(i)}\right]$ is generally not a square matrix. Specifically, $\left[E_{\text{cali}}^{(i)}\right]_{q,k} = E_{\{u_k,v_k\}}^{(i)}(x; l_x = L_q)$, $[y_{\text{cali}}^{(i)}]_q = y_i(x; l_x = L_q)$, and $\{u_k, v_k\}$ is the coordinate of the $k$-th pixel. Notably, the localization of the output field into distinct focal points allows for effective error correction, which is essential for an analog computing device, given the various contributing factors such as fabrication imperfections and optical misalignments.

Fig. 3d shows an example of a single lens illumination. The first row shows the complex illumination field, where each lens is illuminated individually with zero phase. The second and third rows depict the simulated and measured reconstructed fields around $f_2$. The reconstructed complex field reveals that single lens illumination creates a phase grating along the direction from the illuminated signal lens to the reference lens. In this example, the phase difference between the elements of the second row of $T_{ij}$ is $2\pi/8$, corresponding to the hue shift of the center of each reconstructed field. Finally, we show the measured output vector phase $\arg \tilde{y}'$ after error correction on Fig. 3e. Ideally, each value on Fig. 3e should match the complex phase of the $\text{DFT}_8$ matrix. We found that applying error correction minimizes the phase RMSE from 0.30 rad down to 0.08 rad.

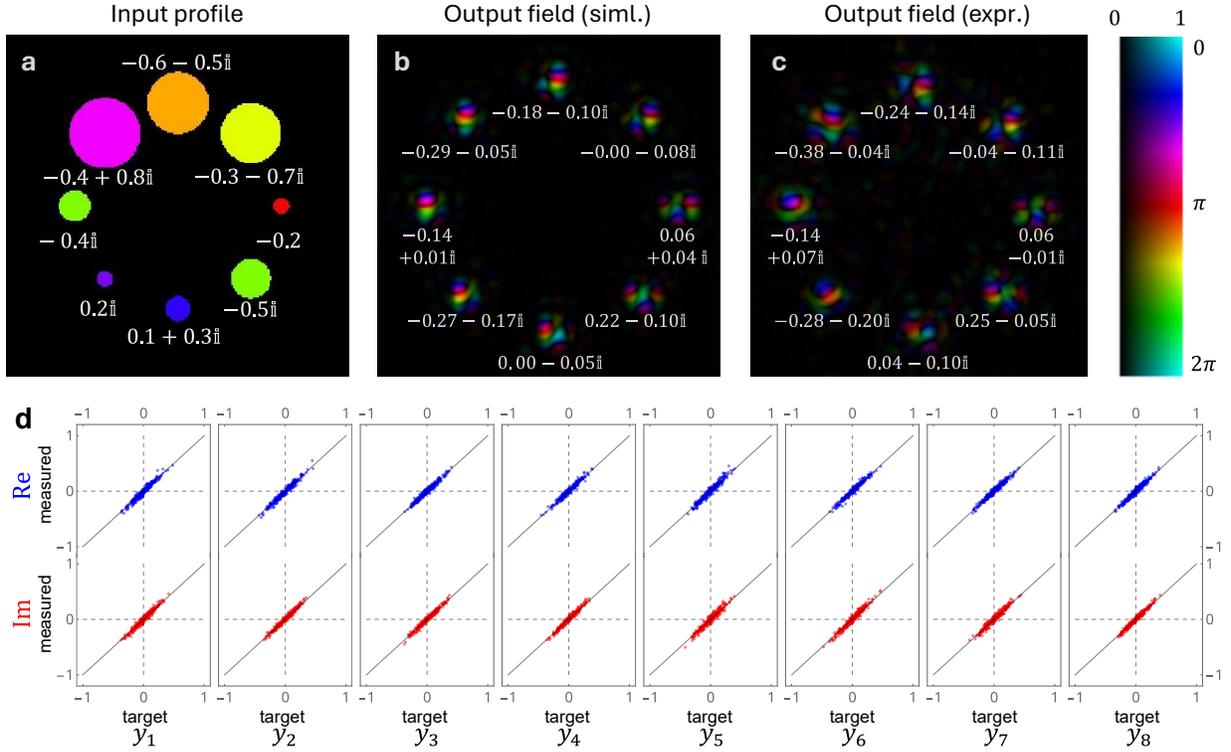

**Figure 4 Performance evaluation on random input test**. **(a)** A typical example of an input vector with a random amplitude and phase. The input amplitude control is proportional to the radius of illumination, while the phase directly set on the SLM. **(b)** Simulated reconstructed field, with numbers showing the target output vector $y = (\text{DFT}_8)\, x$. **(c)** Measured reconstructed field, with numbers showing measured output vector $\tilde{y}$ after error correction. **(d)** Accuracy evaluation for 256 random inputs. x-axes and y-axes are the target and measured value, while blue and red dots correspond to the real and imaginary part. The measurement error is indicated by the vertical deviation of a particular point from the diagonal line.

Fig. 4 presents the performance evaluation of our meta-DFT device on random input vectors. The input vector is generated using a random number generator, with uniform distribution for amplitude $(0, 1)$ and phase $[0, 2\pi)$. Fig. 4a illustrates the complex illumination profile for a typical example of random input vector:

$$x = \{-0.6 - 0.5\mathrm{i}, -0.3 - 0.7\mathrm{i}, -0.2, 0. - 0.5\mathrm{i}, 0.1 + 0.3\mathrm{i}, 0.2\mathrm{i}, -0.4\mathrm{i}, -0.4 + 0.8\mathrm{i}\}$$

For this example, we rounded the real and imaginary part to nearest multiple of 0.1. The input amplitude of each element is proportional to the illumination radius of its corresponding metalens, where an amplitude of 1 indicates full illumination and 0 indicates no illumination.

Fig. 4b shows the simulated reconstructed field, with the numbers showing the target values of DFT $y_i$. The maximum amplitude of any element in the output vector is 1, representing constructive interference when all lenses are fully illuminated with a single input topological charge. Fig. 4c shows the experimentally obtained reconstructed field, with numbers reflecting the measured values derived from decoding each interference pattern. Qualitatively, we observe a strong similarity between the simulated and measured reconstructed fields.

The accuracy of the meta-DFT device is assessed using 256 random input vectors. The real and imaginary parts of the target and measured output vectors are plotted in Fig. 4d. Each plot compares the error-corrected measured output vector $\tilde{y}_i$ (y-axis) against its target value $y_i$ (x-axis). The upper row (blue) displays the real part, while the lower row (red) displays the imaginary part. Each column, from left to right, corresponds to a specific basis $y_i$ extracted from the complex interference pattern at $f_i$. Each point on these plots represents one of the 256 inputs. The measurement error is defined as the vertical deviation of a point from the diagonal line. By only using a single pixel to obtain measurement value, we achieve OFT with linear time complexity $\mathcal{O}(N)$, yielding an RMSE around 0.07 for both real and imaginary part. With error correction enabled, the RMSE is reduced to around 0.03. Notably, while enabling error correction improves accuracy, it also requires additional computational resources that scale with the number of pixels used in the correction process.

**Outlook**

In this work, we present a metasurface platform capable of performing complex-to-complex optical discrete Fourier transforms (DFT). Compared to optical FT based on a 2-f system [37], the metalens array approach is more compact, allows for error correction, and offers versatility for generalizing to arbitrary complex matrices. While we have demonstrated complex-to-complex MVM scheme with $\mathcal{O}(N)$ time complexity, it is important to note that we are still operating in a regime where electronic computers can finish the same tasks at a much faster speed. The current bottleneck arises due to the slow refresh rate of SLM, which is approximately 120Hz – orders of magnitude slower than the clock speed of modern CPUs, around 5GHz.

For a larger sized arbitrary complex MVM, our metalens array approach may prove advantageous. To maintain the equivalent focusing quality when scaling to a larger sized matrix, the device area would need to be scaled linearly with the number of elements of the matrix (Area $\propto MN$ for $M \times N$ matrix). If such a metasurface sample can be fabricated, we estimate that our approach could achieve performance comparable to modern single-core CPUs for matrix size $M \times N \sim 2000 \times 2000$, corresponding to a sample area of around 3cm × 3cm. Additionally, higher-resolution SLMs and imaging systems will be necessary for precise alignment and to resolve fine interference features.

Future work should focus on exploring integrated and programmable meta-devices, dielectric and multi-layer structures, increased clock speed, and advanced error correction scheme. These advancements could enable metasurface-based systems to become practical, high-speed photonic computing applications. Once scaled, such platforms could serve as compact opto-electronic accelerators for existing complex valued neural network models, eliminating the need for inverse designed structures.

We acknowledge support from the Hong Kong RGC through grant no. 16304020, 16306521, 16304524, AoE/P-502/20/, STG3/E-704/23-N and from the Croucher Foundation (CF23SC01).